\documentclass[aps,pra,twocolumn,groupedaddress,showpacs]{revtex4} 

\usepackage{graphicx}
\usepackage{bm}

\begin{document}


\title{Inelastic and elastic  collision rates for triplet  states of ultracold strontium}


\author{A. Traverso, R. Chakraborty, Y. N. Martinez de Escobar,  P. G. Mickelson,  S. B.
Nagel, M. Yan,
 and T. C. Killian}


\affiliation{Rice University, Department of Physics and Astronomy,
Houston, Texas, 77251}




\date{\today}

\begin{abstract}
We report measurement of the inelastic and elastic collision rates
for $^{88}$Sr atoms in the  $(5s5p){^3P_0}$ state in a crossed-beam
optical dipole trap.   This is the first measurement of ultracold
collision properties of a $^3P_0$ level in an alkaline-earth atom or
atom with similar electronic structure. Since the $(5s5p)^3P_0$
state is the lowest level of the triplet manifold, large loss rates
indicate the importance of principle-quantum-number-changing
collisions at short range. We also provide an estimate of the
collisional loss  rates for the $(5s5p){^3P_2}$ state. Large loss
rate coefficients for both states indicate that evaporative cooling
towards quantum degeneracy in these systems is unlikely to be
successful.

\end{abstract}


\maketitle




%

\begin{figure}
  \includegraphics[width=2.25in,clip=true, trim=00 10 230 115, angle=270]{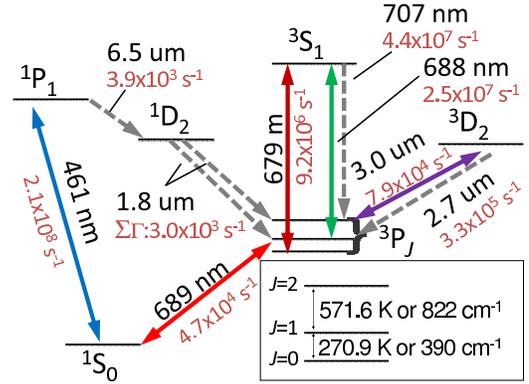}\\ 
  \caption{Strontium atomic levels.
  Decay rates (s$^{-1}$)
   and  excitation wavelengths are given for selected transitions.
  Laser light used for the experiment is indicated by solid lines.
  Dashed lines indicate transitions that are only made due to
  spontaneous decay. All levels with energy below $43,000\,K$ are shown.
  The energy of the $^3S_1$ state is $41780.5\,K$.
  The inset gives the energy splittings
  of the metastable $^3P_J$ levels.
  }\label{levels}
\end{figure}

Metstable $^3P_J$ states of alkaline-earth atoms and atoms with
similar electronic structure (Fig.\ \ref{levels}) display vastly
different optical and ultracold collisional properties compared to
states found in alkali-metal atoms more commonly used in ultracold
atomic physics experiments. The extremely long-lived $^3P_0$ states
in Sr  and Yb serve as the upper levels in neutral-atom optical
frequency standards \cite{ykk08}. The weakly allowed $^1S_0$-$^3P_1$
intercombination transition serves as the basis for powerful
laser-cooling techniques \cite{kii99} and may enable useful optical
tuning of the ground-state scattering length \cite{ctj05}. $^3P_2$
atoms interact through long-range anisotropic interactions
\cite{dpk03,sgr03} that allow magnetic tuning of the interactions
and cause rapid inelastic collisional losses
\cite{ksg03,hhe06,yuh08}. $^3P_J$ states of alkaline-earth atoms
have also been proposed for lattice-based quantum computing
\cite{dca04,dby08}.

Here, we report measurement of the inelastic and elastic collision
rates for $^{88}$Sr atoms in the $(5s5p)^3P_0$ state in a
crossed-beam optical dipole trap. This is the first measurement of
the ultracold collisional properties of a $^3P_0$ state, which is of
great interest because of its role in optical clocks. We also report
an estimate of the collisional loss rates for the $^{88}$Sr
$(5s5p)^3P_2$ state. Large loss-rate coefficients for both states
make efficient evaporative cooling in the ultracold regime unlikely.

Early laser-cooling experiments with Sr \cite{nsl03} and Ca
\cite{hma03} showed that it is straightforward to magnetically trap
metastable $^3P_2$ atoms  through natural decay in a magneto-optical
trap (MOT). This generated interest in the possibility of achieving
quantum degeneracy in this state and motivated calculations that
found novel collisional properties of the metastable $^3P_J$ levels
\cite{dpk03}. Magnetic dipole-dipole and electric
quadrupole-quadrupole interactions between $^3P_2$ atoms produce
anisotropic, long-range potentials with bound states and collisional
rates that can be tuned with magnetic field. Reference \cite{sgr03}
showed that $s$-wave colliding states can be coupled to much higher
partial waves of outgoing channels of other magnetic sublevels even
if the initial state is maximally spin polarized. This leads to
two-body inelastic loss rates in magnetically trapped samples
\cite{ksg03} that are on the order of elastic collision rates,
making efficient evaporative cooling towards quantum degeneracy of
$^3P_2$ atoms in a magnetic trap unlikely. This was confirmed in
experiments with magnetically trapped Ca \cite{hhe06}.


Theory \cite{dpk03,sgr03,ksg03} only considered
Zeeman-sublevel-changing (ZSLC) collisions mediated by long range
interactions, which led to speculation \cite{ksg03,hhe06} that
losses in optical traps might be low enough to allow evaporation.
However, large loss rates were also found in gases of $^3P_2$ Yb
atoms held in an optical dipole trap, which suggested that
fine-structure-changing (FSC) collisions at short-range are also
significant \cite{yuh08}. 

By studying ultracold collisions between $^3P_0$ atoms, which occupy the lowest
level of the fine-structure triplet, we remove the possibility of  ZSLC and FSC
collisions and probe the importance of principle-quantum-number-changing (PQNC)
collisions. For $^3P_0$ atoms, PQNC collisions (also known as energy-pooling
collisions \cite{khg88b}) result in one atom in the ground state. Differences
between $^3P_0$ and $^3P_2$ collisions may also arise because the $^3P_0$ atom
is isotropic and lacks magnetic-dipole and electric-quadrupole moments.
$^{88}$Sr has nuclear spin $I=0$.

At higher temperatures,  de-excitation of Sr$(5s5p)^3P_J$  states
due to collisions with background noble gas atoms \cite{khg88,rse04}
and ground state Sr  \cite{rsf01} has been well-studied.
 Energy-pooling
$(5s5p)^3P_J+(5s5p)^3P_{J'}\rightarrow (5s^2)^1S+(5s6s)^{3,1}S$
collision rates have also been measured in a Sr heat pipe
\cite{khg88b}. All of these studies, as well as ultracold
experiments with $^3P_2$ levels in other atoms \cite{hhe06,yuh08},
worked in a regime where fine-structure changing collisions were
energetically allowed. To our knowledge, we report on the first
study  that isolates
 $^3P_0-{^3P_{0}}$ collisions. 


%
%
%
%
%
%
%
%
%
%



Laser-cooling and trapping aspects of the experiment are described in
\cite{nsl03,mmp08}.
 To obtain higher density and longer sample lifetime, and to trap all electronic states,
  $^1S_0$ atoms are loaded into an optical dipole trap (ODT) of controllable trapdepth as described in
  \cite{mmp08},
yielding equilibrium temperatures of between of 3 and 15\,$\mu$K and peak
densities up to $10^{14}$\,cm$^{-3}$. The potential seen by the atoms is
determined from measured ODT laser properties, well-known atom polarizabilities
\cite{plb08,ykk08}, and measured trap oscillation frequencies \cite{fdw98}.



%


%
%
%
%

\begin{figure}
  \includegraphics[width=2.2in,clip=true, trim=00 30 225 0, angle=270]{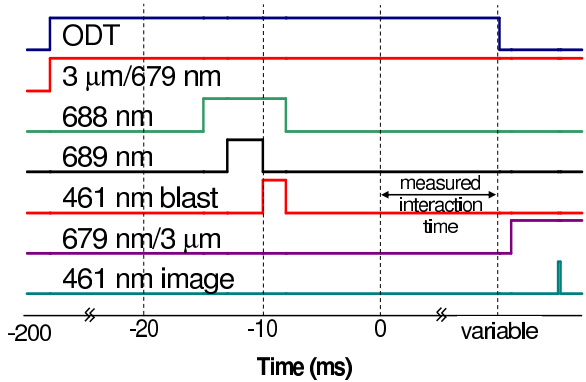}\\ 
  \caption{Timing diagram for measuring metastable collision dynamics (color online). Note the
  breaks in the time axis. Simultaneous application of the 688\,nm
  and 689\,nm lasers populates the metastable triplet levels. A strong 461\,nm beam applied after the initial optical pumping
ensures that no atoms remain in the $^1S_0$ state.
  For measuring $^3P_0$  dynamics and to keep the $^3P_2$ level unpopulated,
  the 3\,$\mu$m laser is on the entire time,
  and the 679\,nm is pulsed on to  repump the
  $^3P_0$ atoms to the ground state.
  For measuring $^3P_2$  dynamics, the timing of these two lasers is interchanged
  as indicated in the timing labeling.
  }\label{figure:timing}
\end{figure}

We then pump atoms via $^1S_0 \rightarrow {^3P_1} \rightarrow {^3S_1}$ followed
by natural decay to the metastable state of interest by applying a 689\,nm
laser for 3 ms while a  688\,nm laser is on (Figs.\ \ref{levels} and
\ref{figure:timing}). Atoms decaying from the $^3S_1$ state to the wrong
metastable state are repumped with a clean-up laser at 3.0\,$\mu$m for
experiments with $^3P_0$ atoms and with a 679\,nm laser for experiments with
$^3P_2$ atoms. The clean-up repumper is turned on during the loading stage of
the ODT and left on for the rest of the experiment. Any atoms remaining in the
ground state are removed from the trap with a 2 ms 461\,nm pulse. We typically
obtain
 $10^6$ metastable atoms at a temperature near 10\,$\mu$K, a
density as high as $10^{13}$\,cm$^{-3}$, and phase-space density as
high as $10^{-3}$.

The zero of time for interaction studies is set after the metastable
atoms have  equilibrated for approximately 10 ms after the 689\,nm
 laser is extinguished. The end of the interaction time
is determined by the extinction of the ODT, after which the atoms
ballistically expand and fall under the influence of gravity. The
density drops rapidly enough that atom-atom interactions cease on a
millisecond time scale.

The number of atoms and sample temperature are determined with
time-of-flight absorption imaging using the $^1S_0$-$^1P_1$
transition. This necessitates repumping the atoms  to the ground
state during the ballistic expansion by applying the 679\,nm laser
in the case of $^3P_0$ atoms and 3.0\,$\mu$m laser in the case of
$^3P_2$ atoms. It is important to release the ODT before repumping
because the repump lasers cause density-dependent light-assisted
losses. Complete repumping requires a few milliseconds.

 Photon recoil during optical pumping  to the metastable state before the interaction time
 and
 then to the
ground state for imaging affects the atomic momentum distribution,
which complicates measurement of the temperature. The
 recoil energy for a red photon is ${\hbar^2 k^2}/(m k_B)=0.5\, \mu$K, and
 from the branching ratios shown in Fig.\
\ref{levels}, each atom is expected to scatter about 5 red photons
during initial excitation and 1-3 photons during repumping. The
repump
 lasers are all aligned  horizontally, so we expect
 several $\mu$K of extra heating along this axis. Recoil from the
$3\,\mu$m photons is insignificant.



 An approximate description of the
 loss of atoms during the interaction time
 can be derived from a local equation for
the evolution of the atomic density, $ \dot{n}=-\beta n^2-\Gamma n
$, which, assuming constant sample temperature, can be integrated
spatially to yield the evolution in atom number
\begin{equation}\label{number}
   N(t)={N_0 \rm{e}^{-\Gamma t} \over 1+
   {N_0 \beta V_2\over \Gamma V_1^2}(1-\rm{e}^{-\Gamma t})}.
\end{equation}
$N_0$  is the number  at the beginning of the interaction time, and the
one-body loss rate, $\Gamma$ is due to background collisions. The effective
volumes weighting one- and two-body processes, indexed by subscript $q$, are
defined by
\begin{equation}
V_{q}(T)={\int \mathrm{d}^3r \, [n(\textbf{r})/n_0]^q},
\end{equation}
where $n(\textbf{r})$ is the spatial density distribution calculated
from the trap potential and atom temperature including effects due
to truncation of the Boltzmann distribution \cite{lrw96}, and $n_0$
is the peak density in the trap.
 The spatial integral extends
over the region contained in the trap.

\begin{figure}
  \includegraphics[width=3.45in,clip=true, trim=30 250 20 00, angle=00]{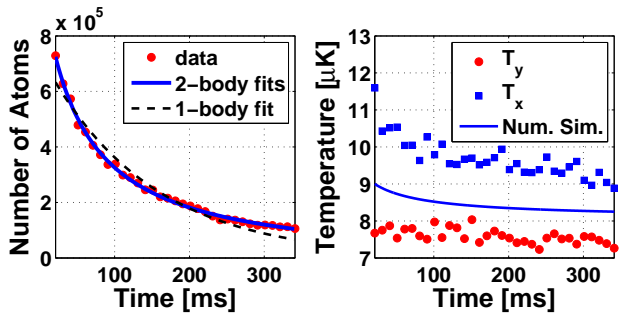}\\ 
  \caption{Left: Number of trapped atoms as a function of time for studies of
  the $^3P_0$ state (color online). Two-body fits allowing for
  one and two-body losses using Eq.\ \ref{number} and the numerical
  simulation described in the text are identical. The statistical
  uncertainties in
  the two-body loss
  rate constant, $\beta=18\pm 3 \times 10^{-18}$\,m$^3$/s, and
 one-body loss rate,
  $\Gamma=1.4\pm 0.9$\,s$^{-1}$ are dominated by correlation between
  the parameters, but $\beta$ is determined much better than
  systematic uncertainties described in the text and is the dominant
  loss mechanism. The
  one-body fit sets $\beta=0$ and is not a good description of the
  data. The observed value for $\Gamma$ is consistent with the decay rate
  for $^1S_0$ atoms, which is dominated by background gas collisions.
  Right: Temperature evolution of the trapped sample.
  The fit is from the numerical
  simulation.
  The ODT trap depth resulting from symmetric saddle-points located approximately
  along
  the horizontal axes ($x$ and $z$) is  $U_{trap}/k_B=20\, \mu$K.
  The potential barrier for escape along the vertical ($y$)
  direction
  is $30 \mu$K.
  }\label{Figure:twobodyproof}
\end{figure}

Figure \ref{Figure:twobodyproof} shows representative data for
temperature and number evolution for atoms in the $^3P_0$ state. The
horizontal axis ($x$) is about 2\,$\mu$K hotter than the vertical,
which we attribute to photon recoil heating during optical pumping.
The temperature change during the interaction time and the $x-y$
temperature difference are both small, so an assumption of constant
temperature is a reasonable approximation, and Eq. \ref{number} fits
the data well. The sample temperature used to calculate effective
volumes is taken as the average of $T_x$ and $T_y$. A fit neglecting
two-body decay reproduces the data poorly, confirming the importance
of two-body effects.

\begin{figure}
  \includegraphics[width=3.5in,clip=true, trim=10 250 0 0, angle=00]{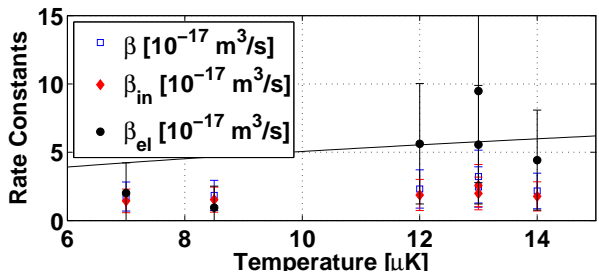}\\ 
  \caption{Two-body collision rate constant measurements for the $^3P_0$ state (color
  online). The solid line is $\beta_{el}=\sigma \overline{v} \sqrt{2}$, where $\sigma=8\pi a^2$,
  $|a|=100\,a_0$,
    and $\overline{v}=\sqrt{8 k_B T/\pi m}$.
  }\label{Figure:beta}
\end{figure}

Values of $\beta$ derived from a series of decay curves are shown in
Fig.\ \ref{Figure:beta}. The dominant uncertainties in $\beta$ are
systematic and reflected in the error bars. Uncertainty in
temperature of $2\,\mu$K only contributes a 10\% uncertainty in
$\beta$, while uncertainty in the trap oscillation frequencies at
the 20\% level causes a 60\% uncertainty. At our level of
  uncertainty, we observe no temperature dependence and find
  $\beta=(2.3 \pm 1.4) \times10^{-17}$\,m$^3$/s.

As pointed out in \cite{ddo04}, both inelastic-collisional loss and
evaporation due to elastic collisions are two-body loss processes
contributing to $\beta=\beta_{in}+f\beta_{el}$, where $\beta_{in}$
is the inelastic-collision-loss rate constant, $\beta_{el}$ is the
elastic-collision rate constant, and $f$ is the fraction of elastic
collisions resulting in an evaporated atom. The
 rate constants can be expressed as
 $ \beta_{in} =  {\beta}/{(1+ f\gamma)}$ and $\beta_{el}
= {\gamma \beta}/{(1+ f\gamma)}$,
 where the ratio of rate constants is
\begin{equation}\label{equation:elasticinelasticratio}
    \gamma\equiv\frac{\beta_{el}}{\beta_{in}}
=\frac{\bar{E}-\bar{E}_{in}}{f(\bar{E}_{el}-\bar{E})}.
\end{equation}
 $\bar{E}$ is the average energy of the trapped atoms,
$\bar{E}_{in}$ is the average energy of atoms lost due to inelastic
collisions, and $\bar{E}_{el}$ is the average energy of atoms lost
due to evaporation. Parameters on the r.h.s of Eq.\
\ref{equation:elasticinelasticratio}
 can be calculated from
 the trapping potential and equilibrium atom temperature.
  In our case,
these quantities must be evaluated numerically because the potential is not
amenable to analytic solutions and truncation of the Boltzmann distribution is
significant \cite{lrw96}. For the various conditions studied here the ranges of
parameters are $0.7<f<0.3$ and $0.6<\gamma< 3.7$, and we obtain
$\beta_{in}=(1.9\pm 1.2)\times 10^{-17}$\,m$^3$/s (Fig.\ \ref{Figure:beta}).
The data suggests that $\beta_{el}$ increases slightly with temperature, as one
would expect since $\beta_{el}=\sigma \overline{v} \sqrt{2}$, where $\sigma$ is
the elastic scattering cross section and $\overline{v}$ is the mean thermal
velocity.
 The error bars in
these measurements are dominated by the trap uncertainties. Our treatment
assumes ergodicty, which is reasonable given the small difference between $T_x$
and $T_y$.


To check that the assumption of an equilibrium temperature does not introduce
significant bias, we adapted a numerical model of the dynamics of atoms in an
ODT \cite{cfs06} designed for a system with high $\eta \equiv U_{trap}/k_B T$
to our low-$\eta$ situation by incorporating the evaporation treatment of
\cite{lrw96} and numerically calculating all effective volumes and related
integrals using the actual trap shape and a truncated momentum distribution.
Fits to both number and temperature evolution (Fig.\ \ref{Figure:twobodyproof})
yielded the same values of $\beta_{in}$ and $\beta_{el}$ as above. Assuming
pure $s$-wave elastic collisions, this simulation also allows us to provide an
estimate of the magnitude of the $^3P_0$ scattering length $|a|=100\pm 50\,
a_0$.


 Measurement of the number and temperature evolution for atoms in the $^3P_2$
 state, together with numerical simulations
 assuming an equal trap depth for all magnetic
 sublevels, constrain
   $\beta_{in}=(1.3 \pm 0.7 \pm 0.7) \times10^{-16}$\,m$^3$/s reasonably well.
   The first
  uncertainty comes from the fit, and the second uncertainty is
  systematic from knowledge of the trap potential.
   Several factors that increase uncertainty in determination of $\beta$
 in this case must be addressed.
 The temperature measurement along $x$ is distorted by
 a high density of atoms evaporating into the arms of the
  crossed-dipole trap. 
 However, the numerical simulation and elastic-collision rate
 implied by the low
 $T_y$  indicate that the  temperature change and lack of equilibrium
 is not significantly worse than in $^3P_0$.
 The different magnetic sublevels have different AC stark shifts and
  experience
  optical trap depths that vary by almost a factor of two.
  This complicates modeling of the
   system  and introduces
   ZSLC-collisional heating and loss processes. But this would only
   increase $\beta$, and we take the observed  value as an upper limit for the
  rate constant for FSC and PQNC collisional loss in this state.
  The scattering length extracted from the simulation, $a\approx600\,a_0$, is less reliable because of its
   sensitivity to the trap depth.

  The
$\beta_{in}$ for Sr $^3P_0$, which can only reflect PQNC collisions, is a
factor of two larger than $\beta_{in}$ for optically trapped Yb $^3P_2$
\cite{yuh08} and  a factor of 5 less than in Sr $^3P_2$. The $^3P_2$ states are
sensitive to PQNC \textit{and}
 FSC processes.
ZSLC collisional loss rates in magnetically trapped $^3P_2$ are one order of
magnitude greater \cite{ksg03}, as found in magnetically trapped Ca
\cite{hhe06}.

 One expects higher PQNC collision rates for
$^3P_2$ collisions than for $^3P_0$ collisions because the process
${^3P_J}+{^3P_{J}}\rightarrow
{(5s^2){^1S_0}}+{(5s6s){^{3}S_1}}+\Delta E$ is allowed for $J=2$
($\Delta E/k_B=1091\,K$) and energetically suppressed for $J=0$
($\Delta E/k_B=-581\,K$). (See Fig.\ \ref{levels}.) In light of the
observed values of $\beta_{in}$ this implies that PQNC collisions
make a significant contribution to the inelastic collision rate
 for ultracold $^3P_2$ states. And as predicted
\cite{ksg03}, ZSLC collisions should be much more rapid.

In conclusion, this work has highlighted the important role of PQNC
collisions in ultracold metastable triplet levels in
two-valence-electron atoms. It has also provided the first
measurement of the inelastic and elastic collision properties for
ultracold atoms in a $^3P_0$ state and an estimate of collisional
properties for $^3P_2$ Sr atoms. These levels are of significant
current interest in applications such as atomic clocks and in
fundamental studies in ultracold atomic physics. The techniques
developed here demonstrate the production of a cold and dense sample
of  $^3P_0$ Sr atoms, which is a prerequisite for photoassociative
spectroscopy of atoms in this state and may allow determination of
dipole matrix elements and black-body radiation shifts that limit
the accuracy of optical clocks \cite{pde06PRA,ykk08}.


This research was supported by the Welch Foundation (C-1579), National Science
Foundation (PHY-0555639), and the Keck Foundation. We thank D. Comparat for
sharing the numerical code for simulating evaporative cooling.

\end{document}